\documentclass[10pt, preprint]{aastex} 

\newcommand{\BII}{B$\;${\small\rm II}\relax}
\newcommand{\OI}{O$\;${\small\rm I}\relax}
\newcommand{\CI}{C$\;${\small\rm I}\relax}

\newcommand{\htwo}{H$_2$}

\newcommand{\nav}{$N_a (v)$}

\newcommand{\e}[1]{10^{#1}}
\newcommand{\hd}{HD~}
\newcommand{\err}[2]{\ensuremath{^{+ #1}_{- #2}}}

\newcommand{\kms}{km~s$^{-1}$\relax}

\newcommand{\avedens}{$\langle n_{\rm H} \rangle$}
\newcommand{\fhtwo}{$f({\rm H_2})$}

\newcommand{\asco}{$\alpha$~Sco B}
\newcommand{\zoph}{$\zeta$~Oph}

\newcommand{\stis}{STIS}

\begin{document}

\title{The Abundance of Interstellar Boron\altaffilmark{1}}

\altaffiltext{1}{Based on observations made with the NASA/ESA Hubble Space
Telescope, obtained from the data archive at the Space Telescope
Science Institute. STScI is operated bythe Association of Universities
for Research in Astronomy, Inc. under the NASA contract NAS 5-26555. }

\author{J. Christopher Howk\altaffilmark{2}, 
Kenneth R. Sembach\altaffilmark{2}, \& Blair D. Savage\altaffilmark{3}}

\altaffiltext{2}{Department of Physics \& Astronomy, 
The Johns Hopkins University, Baltimore, MD 21218, \begin{sloppypar}
howk@pha.jhu.edu, sembach@pha.jhu.edu \end{sloppypar}}

\altaffiltext{3}{Department of Astronomy, University of Wisconsin-Madison, 
        Madison, WI 53706, savage@astro.wisc.edu}


\slugcomment{Accepted for publication in {\em The Astrophysical Journal}.}

\begin{abstract}

We use new Space Telescope Imaging Spectrograph (\protect\stis) and
archival Goddard High Resolution Spectrograph (GHRS) observations to
study interstellar \protect\ion{B}{2} $\lambda1362$ and
\protect\ion{O}{1} $\lambda1355$ absorption along seven sightlines.  Our new
column density measurements, combined with measurements of four
sightlines from the literature, allow us to study the relative B/O
abundances over a wide range of interstellar environments.  We measure
sightline-integrated relative gas-phase abundances in the range $[{\rm
B/O}] = -1.00$ to $-0.17$, and our data show the B/O abundances are
anticorrelated with average sightline densities over the range $\log
\langle n_{\rm H} \rangle \approx -1.3$ to $+0.7$.  Detailed comparisons 
of the \protect\ion{B}{2} and \protect\ion{O}{1} line shapes show that
the B/O ratio is significantly higher in warm interstellar clouds than
in cool clouds.  These results are consistent with the incorporation
of boron into dust grains in the diffuse ISM.  Since boron is likely
incorporated into grains, we derive a lower limit to the present-day
total (gas+dust) interstellar boron abundance of ${\rm B/H} \ga
(2.5\pm0.9)\times \e{-10}$.  The effects of dust depletion and
ionization differences from element to element will make it very
difficult to reliably determine $^{11}$B/$^{10}$B along most
interstellar sightlines.

\end{abstract}



\section{Introduction}

Production of the light element boron in standard big bang
nucleosynthesis models is negligible (see, e.g., Pagel 1997).  The
chemical evolution of boron is determined by its production through
cosmic ray spallation of interstellar material (Reeves, Fowler, \&
Hoyle 1970; Meneguzzi, Audouze, \& Reeves 1971) and/or
neutrino-induced spallation in type II supernovae (Woosley et al.
1990) and its destruction via astration.  The evolution of the cosmic
boron abundance therefore reflects the cosmic ray flux and/or
supernova rate history in the Milky Way.  Measurements of stellar
boron abundances have revealed a trend of increasing B/H over time (as
expected), from ${\rm B/H} \sim 2 \times\e{-12}$ in very metal-poor
halo stars (Edvardsson et al. 1994; Duncan et al. 1997) to the much
higher Orion association (${\rm B/H} \approx [3-6]\times
\e{-10}$; Cunha et al. 1997) and solar system (meteoritic) abundances
(${\rm B/H} = [7.6\pm0.7] \times\e{-10}$; Anders \& Grevesse 1989).
Because there is evidence that the solar system may be enhanced in
metals such as oxygen relative to the local interstellar medium (ISM;
e.g., Meyer et al. 1998), young B stars (e.g., Gies \& Lambert 1992;
Kilian-Montenbruck et al. 1994) and \protect\ion{H}{2} regions (e.g.,
Peimbert, Torres-Peimbert, \& Dufour 1993), it is unclear that the
solar system B/H is the best fiducial point for the present-day boron
abundance in studies of Galactic chemical evolution.  The boron
abundance of the present-day ISM may be more appropriate for such
studies.  The differences between the solar system and the ISM could
yield important information on the recent chemical evolution of the
solar neighborhood.

Here we present new high-resolution observations of interstellar \BII\
$\lambda1362.461$ and \OI\ $\lambda1355.598$ absorption towards the
stars \hd 104705, \hd 121968, \hd 177989, \hd 218915, and \hd 303308
taken with \protect\stis\ on-board the {\it Hubble Space Telescope
(HST)}.  We also present analyses of the B/O abundances along the
$\alpha$ Scorpii B and $\zeta$ Ophiuchi sightlines using archival
GHRS datasets.  We have more than doubled the measurements
of interstellar boron, and the available data now probe a wide range
of diffuse ISM environments.  After discussing our observations in \S
\ref{sec:observations}, we will show in \S \ref{sec:results} that the
present-day interstellar B/H ratio seems to be dominated by the
incorporation of boron into dust grains.  Thus the interstellar
gas-phase boron abundance provides only a lower limit to the total
(gas+dust) interstellar boron abundance.

\section{Observations and Reductions}
\label{sec:observations}

The \protect\stis\ data used here were obtained as part of our
\protect\stis\ GO program (\#7270), with the exception of the \hd
303308 data, which were obtained from the \protect\stis\ archive.  In all
cases the observations employed the far-ultraviolet MAMA detector and
the E140H grating with the light from the star passing through the
$0\farcs2\times0\farcs09$ aperture.  The resolution of these data is
$\Delta v \sim 2.7$ \kms\ (FWHM).  The data were extracted and the
backgrounds estimated as described by Howk \& Sembach (2000).  For
spectral regions covered in multiple orders (or observations), we have
coadded the flux-calibrated data and weighted the contribution of each
individual spectrum by the inverse-square of its error vector.

The archival, post-COSTAR GHRS data used in this work were
reduced as discussed by Howk, Savage, \& Fabian (1999).  The \asco\
data were obtained through the large science aperture, while the
\zoph\ data were obtained through the small science aperture.  Both
datasets employed the Ech-A grating, yielding a resolution of $\Delta
v \sim3.5$ \kms\ (FWHM).

Figure \ref{fig:spectra} shows the \protect\stis\ spectra and Table
\ref{tab:eqwidths} gives the measured equivalent widths,
$W_\lambda$, for the new \protect\stis\ and GHRS measurements.  The
continua were estimated using low-order Legendre polynomial fits to
regions free of interstellar lines, and our $1\sigma$ error estimates
include continuum placement uncertainties and the effects of 2\%\
zero-point uncertainties for both spectrographs (Sembach \& Savage
1992).  

Federman et al. (1993) and Lambert et al. (1998) have studied the
\BII\ absorption along the \zoph\ sightline using the GHRS with 
discrepant results.  Federman et al. (1993) derive $W_\lambda =
1.78\pm0.29$ m\AA\ using the G160M grating, while Lambert et
al. (1998) derive $W_\lambda = 0.6\pm0.2$ m\AA\ using the Ech-A
grating.  Given the differences in these studies, we have reanalyzed
both GHRS datasets.  We derive an equivalent width using the Ech-A of
$W_\lambda = 0.79\pm0.17$ m\AA, within $1\sigma$ of the Lambert et
al. result.  We do not find clear evidence for \BII\ absorption in the
small science aperture G160M grating data to a $3\sigma$ limiting
equivalent width of $W_\lambda < 1.3$ m\AA, consistent with our Ech-A
results.  However, the continuum placement uncertainties are large in
this region of the spectrum when using the intermediate-resolution
G160M grating.  Alternative continuum placements could push this
limiting equivalent width as high as 2.1 m\AA.  We believe that the
Ech-A results are the most reliable for this sightline and adopt
$W_\lambda = 0.79\pm0.17$ m\AA\ for the \protect\BII\ absorption
towards \protect\zoph.

The sightline towards \protect\asco\ also deserves comment.  The
sightline to this star passes through the stellar wind of the M1.5 Ib
primary (Antares), which lies $2\farcs9$ from the \asco\ sightline
(van der Hucht, Bernat, \& Kondo 1980; Bernat 1982; Cardelli 1984).
It is therefore possible that the absorption lines in the spectra of
\asco\ probe the wind of the primary star.  We do not believe that the 
\protect\OI\ and \protect\BII\ absorption seen in the archival GHRS 
spectra are caused by the stellar wind.  These species show two blends
of material centered at $v_{\rm LSR} = -4.1$ and +3.5 \kms.  Material
associated with the wind is centered near $v_{\rm LSR} = -18$ \kms\
(van der Hucht et al. 1980).  The archival GHRS dataset contains good
observations of lines that trace only the outflowing stellar wind,
notably \ion{Ti}{2}, \ion{Ti}{2}$^*$, \ion{Ti}{2}$^{**}$, and
\ion{S}{1}$^{**}$.  These lines are clearly shifted with respect to
the \protect\OI\ and \protect\BII\ absorption.  They are centered near
$v_{\rm LSR} = -18$ \kms, often with wings extending towards more
negative velocities.  There is no evidence for wind material at the
velocities of the \protect\OI\ and \protect\BII\ absorption.
Therefore, we believe the absorption line measurements presented in
Table \ref{tab:eqwidths} trace the ISM in this direction rather than
the stellar wind of Antares.\footnote{We also note that the column
density of \protect\OI\ derived below is comparable to that of other
stars in this region of sky with similar distances (Meyer et
al. 1998).}

\section{Results and Discussion}
\label{sec:results}

Table \ref{tab:abundances} gives the derived column densities of \OI\
and \BII\ for the stars listed in Table \ref{tab:eqwidths}.  Also
given are the total hydrogen column densities and the normalized
relative gas-phase abundances [B/O]\footnote{We define 
$[{\rm B/O}] \equiv \log N(\mbox{\protect\ion{B}{2}}) / 
N(\mbox{\protect\ion{O}{1}}) - \log ({\rm B/O})_\odot$ 
and assume a meteoritic abundance $\log ({\rm B/O})_\odot = -5.99$
(Anders \& Grevesse 1989).}
for these sightlines.  We also compile all measurements of
interstellar \BII\ using {\em HST} data from the literature (with
references given in the table).

The new \OI\ and \BII\ column densities were derived by integrating
the apparent optical depth profiles (Savage \& Sembach 1991) of each
line.  In a few cases we have deemed it necessary to apply moderate
saturation corrections to the \protect\OI\ column densities.  We have
tested for saturation problems and derived the necessary corrections
by applying a single-component curve of growth to the measured
\OI\ equivalent widths (Table \ref{tab:eqwidths}), adopting $b$-values 
derived from a curve-of-growth fit to several \CI\ lines for each
sightline.  The \OI\ absorption is dominated by narrow components that
are also strong in species that trace dense clouds such as \CI,
\ion{S}{1}, \ion{Cl}{1}, and CO.  While \CI\ and \protect\OI\ need not
be coexistent, the emperical association of \protect\CI\ and
\protect\OI\ absorption, particularly in the components where the
saturation is likely to be greatest, gives us confidence that
\protect\CI\ saturation effects provide a suitable means for
understanding the \protect\OI\ saturation along these sightlines.  The
\protect\OI\ apparent column densities we obtained for four of the 
sightlines from Table \ref{tab:eqwidths} required moderate saturation
corrections of +0.06 to +0.08 dex based on the curve-of-growth fits,
and these have been noted in Table \ref{tab:abundances}.

Figure \ref{fig:b2o} shows the sightline-integrated gas-phase
abundances [B/O] as a function of average line of sight hydrogen
densities, \avedens\footnote{Defined $\langle n_{\rm H} \rangle \equiv
N({\rm H})/d_*$, where $d_*$ is the distance to the star from Table
\ref{tab:abundances}.}, for the 11 sightlines from Table 
\ref{tab:abundances}.  Where no \htwo\ column density measurements 
exist we adopt a generous $+0.25$ dex uncertainty in the total
hydrogen column, $N({\rm H})$, to account for the unknown contribution
from molecular material.  The bottom panel of Figure \ref{fig:b2o}
shows [O/H] versus \avedens\ for the same sightlines.  The dashed line
shows the average value of [O/H]($=-0.34\pm0.02$) from the sightlines
studied by Meyer et al. (1998), several of which are included in this
work.\footnote{We adopt the \OI\ $f$-value suggested by Morton (2000),
which implies a $+0.03$ dex correction to the \OI\ column densities in
Meyer et al. (1998).}

Figure \ref{fig:b2o} shows a clear trend of decreasing [B/O] abundance
with increasing \avedens.  Such a trend is often observed for species
incorporated into interstellar dust grains (cf., Jenkins 1987).  The
general decrease in gas-phase abundance with increasing \avedens\
reflects the mixture of cold and warm diffuse clouds along the line of
sight (Jenkins, Savage, \& Spitzer 1986), where the cold clouds
exhibit a greater incorporation of elements into interstellar grains.

All of the STIS targets in Table \ref{tab:abundances} have distances
$d_* > 1000$ pc, while the GHRS targets are all at $d_* < 500$ pc.  It
is conceivable that the gas being probed by the STIS data is also
significantly more distant than that being probed by the GHRS data.
In this case, some of the behavior seen in Figure \ref{fig:b2o} could
be caused by abundance gradients.  There are several reasons to
believe this is not the case.  First, of the 11 stars in our sample
(Table \ref{tab:abundances}), 8 have galactocentric distances between
7.5 and 9.5 kpc, i.e., they lie within 1 kpc of the solar circle.
Within this 2 kpc in galactocentric distance, the [B/O] abundance
varies from $-0.31\err{0.10}{0.12}$ to $-1.00\err{0.10}{0.12}$, i.e.,
the abundance varies by a factor of 5.  Thus, most of the trend seen
in Figure \ref{fig:b2o} occurs within 1 kpc galactocentric distance of
the solar circle and cannot be caused by large-scale abundance
gradients.

\begin{figure}
\epsscale{0.95}
\plotone{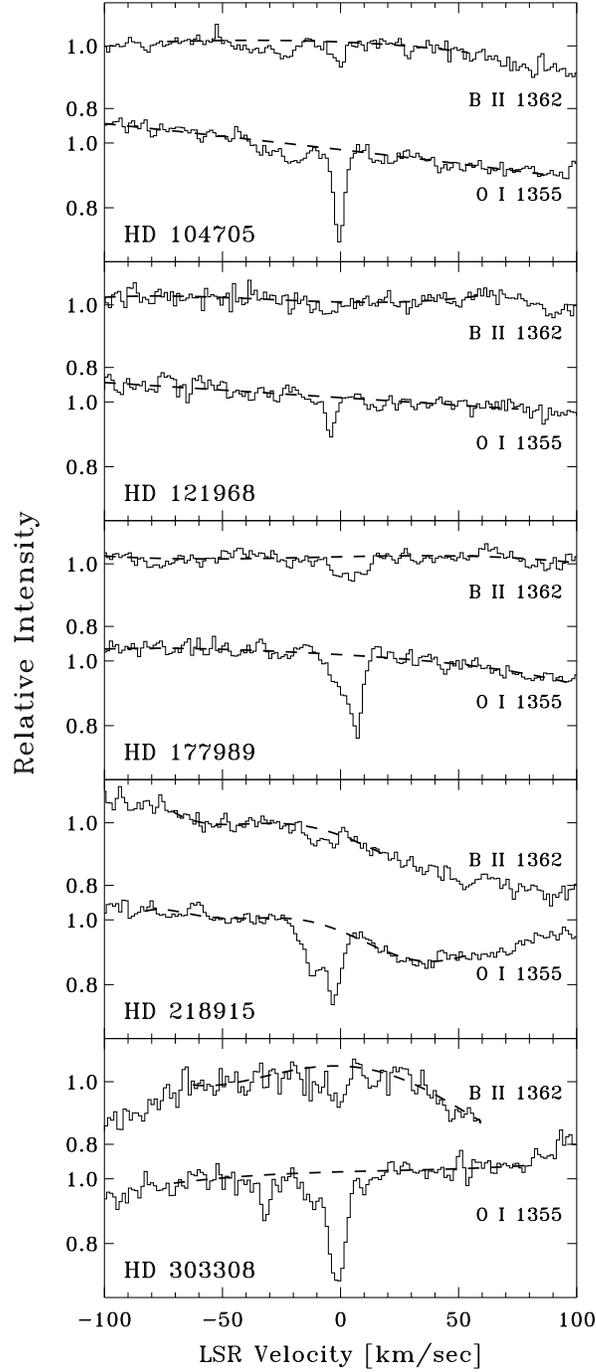}
\caption{STIS absorption line spectra of \protect\ion{B}{2} 
$\lambda1362$ and \protect\ion{O}{1} $\lambda1355$ for five extended
sightlines ($d_* \ga 2.5$ kpc) through the Milky Way.  Our fits to the
local stellar continua are shown as the dashed lines.  The velocity
ranges over which we have integrated these spectra are given in Table
\protect\ref{tab:eqwidths}.
\label{fig:spectra}}
\end{figure}

\begin{figure}
\plotone{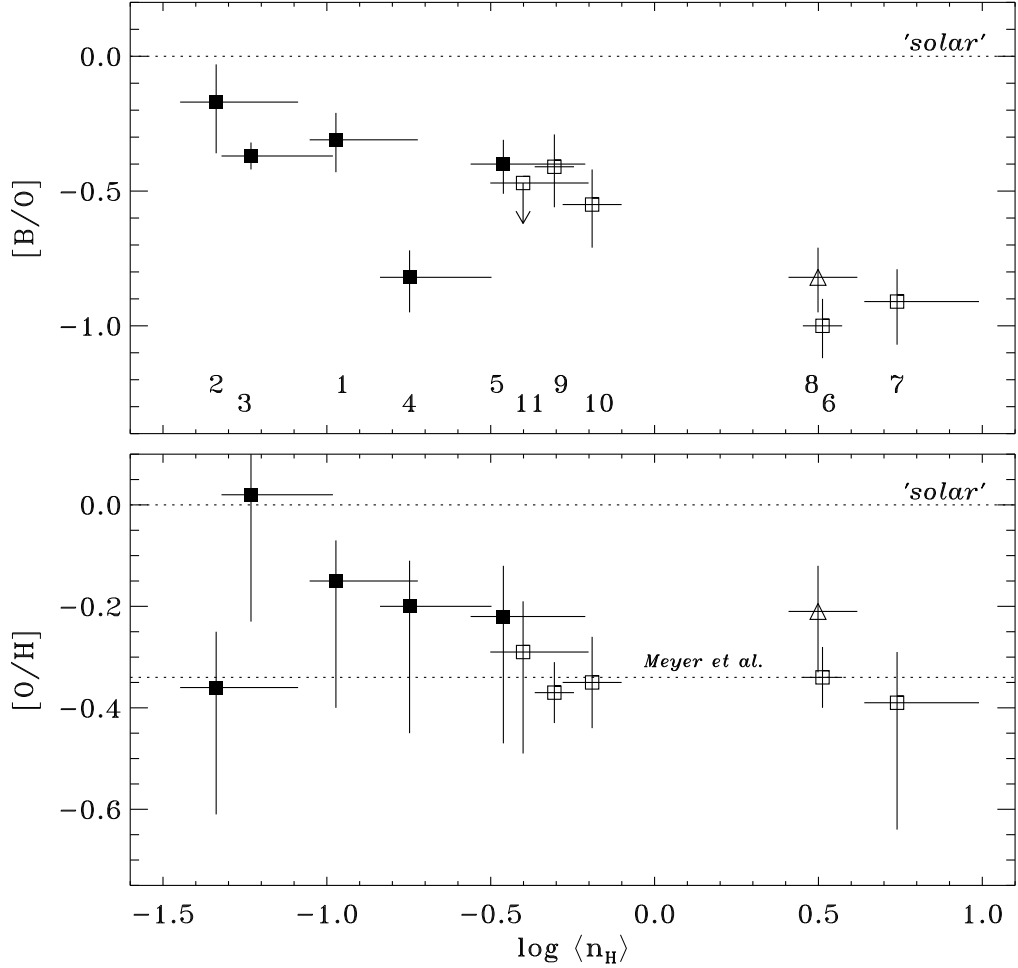}
\caption{Measurements of sightline-integrated [B/O] 
({\it top panel}) and [O/H] ({\it bottom panel}) abundances as a
function of the average sightline hydrogen density, $\langle n_{\rm H}
\rangle(\equiv N({\rm H})/d_*)$.  The numbers in the top panel identify 
each sightline according to the ID in Table
\protect\ref{tab:abundances}.  The filled squares are new measurements
derived from STIS observations of
\protect\ion{O}{1} and \protect\ion{B}{2}, and the open squares 
are GHRS measurements taken from the literature or from an analysis of
data in the {\em HST} archive.  The open triangle is a GHRS
\protect\ion{B}{2} measurement coupled with a {\it Copernicus}
\protect\ion{O}{1} measurement.  The dashed line in the lower panel
shows the average [O/H] value from the measurements of Meyer et
al. (1998) after correcting for our adopted \protect\ion{O}{1}
$f$-value.  The [O/H] measurements with the 0.25 dex downward error
bars represent sightlines for which $N({\rm H}_2)$ has not been
measured, and $N({\rm H})$ is assumed to be $N(\mbox{\ion{H}{1}})$.
\label{fig:b2o}}
\end{figure}

Second, there is evidence that the absorption along all of the
sightlines actually arises within the first 1-2 kpc.  Most of the
absorption seen in Figure \ref{fig:spectra} is at velocities that are
consistent with nearby material, and given that typical cloud-to-cloud
velocity dispersions are of order $\sigma \sim 8$ \kms\ (Sembach \&
Danks 1994), most of the gas is likely very local.  For example,
although the extended STIS sightlines towards \hd 177989 and \hd
218915 pass over known spiral arms with prominent absorption in other
species (e.g., \ion{Mg}{2}, \ion{Mn}{2}, \ion{Ni}{2},
\ion{Cu}{2}, and \ion{Ge}{2}), there is no evidence for any absorption 
in the weak \OI\ and \BII\ lines from those distant structures.  The
nearest arms probed along these sightlines are the Perseus arm seen
towards \hd 218915 at a distance of $\sim2.5$ kpc, and the Sagittarius
arm towards \hd 177989, which is likely at a distance of $\sim1.8$
kpc.  The Perseus and Sagittarius arms are seen seen in absorption in
other species at $v_{\rm LSR} \approx -45$ \kms\ and $v_{\rm LSR}
\approx +18$ \kms\ towards \hd 218915 and \hd 177989, respectively.

\begin{figure}
\plotone{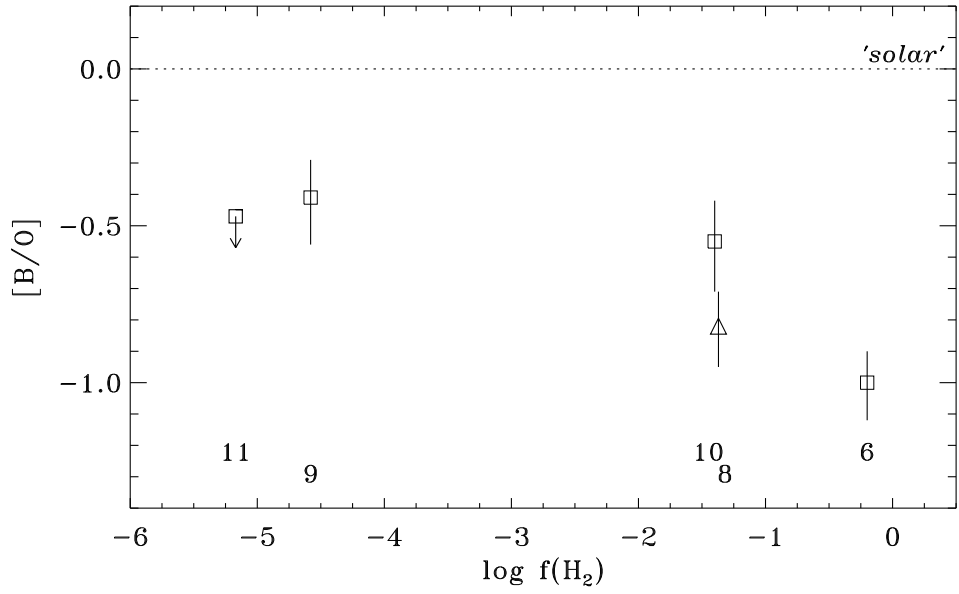}
\caption{Measurements of sightline-integrated [B/O] abundances as a function 
of logarithmic fraction of hydrogen in molecular form, $f({\rm H_2})
\equiv 2N({\rm H_2})/N({\rm H})$, for the sightlines with measured 
$N({\rm H_2})$.  The sightline numbering and symbol definitions are
the same as in Figure \protect\ref{fig:b2o}.  
\label{fig:fh2}}
\end{figure}

The relatively local origin of the gas towards \hd 218915 may explain
why it does not fit well the general trend with average sightline
density seen in Figure \ref{fig:b2o} (point 4 in this figure).  One
would prefer to use a more physically-meaningful measure of physical
conditions such as the fraction of hydrogen in molecular form, $f({\rm
H_2}) \equiv 2N({\rm H_2})/N({\rm H})$.  Figure \ref{fig:fh2} shows
the [B/O] abundances versus \fhtwo\ for those sightlines with $N({\rm
H_2})$ measurements.  This diagram is sparsely populated, but there
seems to be a slight trend of increasing [B/O] abundance with
decreasing \fhtwo.  For this diagram to be truly useful, however, more
$N({\rm H_2})$ measurements are needed.  The {\em Far Ultraviolet
Spectroscopic Explorer} will soon provide molecular hydrogen column
densities for a large number of sightlines towards distant stars,
making it possible to fill in the missing points in Figure
\ref{fig:fh2}.

The trend seen in Figure \ref{fig:b2o}, and the fact that the more
distant stars studied by STIS show higher [B/O] abundances, is likely
caused by the heights of these stars above the plane of the Galaxy.
Because the stars studied by STIS are more distant, they generally lie
at larger distances from the Galactic plane than do the stars studied
by the GHRS.  Thus, the sightlines probed by our STIS measurements
probe lower density regions, on average, than the sightlines probed by
the GHRS measurements.  The well-documented trend of higher gas-phase
abundances of most elements in lower density regions suggests that we
should expect the segregation of of STIS and GHRS measurements
observed in Figure \ref{fig:b2o}.

\begin{figure}
\plotone{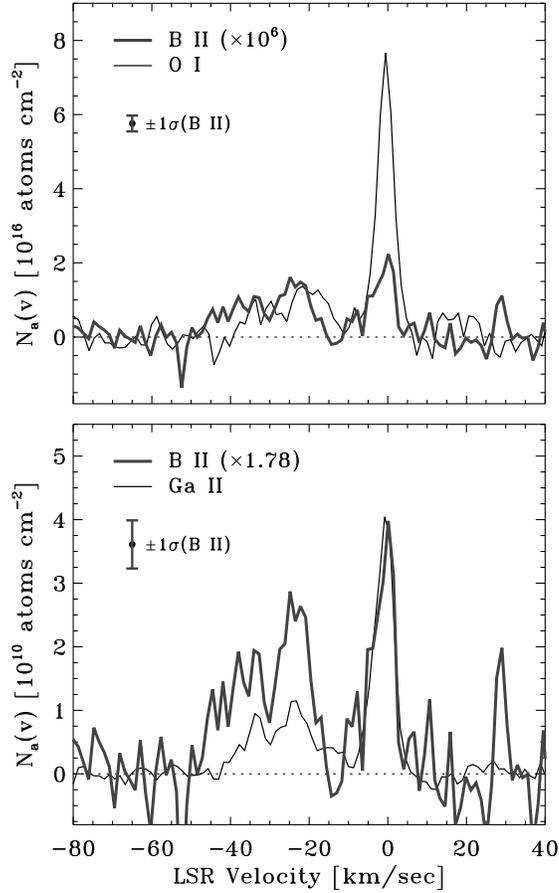}
\caption{Apparent column density, or $N_a(v)$, profile for 
\protect\ion{B}{2} $\lambda1362$ plotted with those of \protect\ion{O}{1} 
$\lambda1355$ ({\it top}), and \protect\ion{Ga}{2} $\lambda1414$ ({\it
bottom}) for the line of sight to \protect\hd 104705.  The
\protect\ion{B}{2} profiles have been scaled by the appropriate solar
system abundances (Anders \& Grevesse 1989) of O/B ($10^6$) and Ga/B
($1.78$).  The narrow component near $v_{\rm LSR} = 0$ km s$^{-1}$
shows sub-solar B/O ratios, while the warm gas between $v_{\rm LSR} =
-40$ and $-15$ km s$^{-1}$ exhibits almost solar B/O.  The narrow
component also shows solar B/Ga abundance, while the negative velocity
material has super-solar relative B/Ga abundances.  This behavior
suggests boron is incorporated into dust and is more readily-stripped
from the grains than gallium. A typical $\pm 1 \sigma$ error bar is
shown for the
\ion{B}{2} profile.
\label{fig:nav}}
\end{figure}

While the average sightline [B/O] values show a clear trend with
\avedens, an imperfect measure of sightline properties, there is also
evidence within the observed line profiles for variation of [B/O] with
the physical properties of the absorbing material.  Several of the
sightlines displayed in Figure \ref{fig:spectra} show evidence for
dense clouds with lower B/O ratios than warm clouds along the same
line of sight, in qualitative agreement with the Jenkins et al. (1986)
model of integrated sightline properties.  Figure
\ref{fig:nav} shows the apparent column density (Savage \& Sembach
1991), or $N_a(v)$, profile of \BII\ $\lambda 1362$ towards
\hd 104705 with the corresponding \nav\ profiles of \OI\ $\lambda
1355$ and \protect\ion{Ga}{2} $\lambda1414$.  This sightline exhibits
a narrow, cold component centered at $v_{\rm LSR}=0$ \kms, which is
prominent in species such as \protect\ion{O}{1}, \protect\ion{S}{1},
\protect\ion{Cl}{1}, and CO, as well as a blend of warmer components
between $v_{\rm LSR}=-40$ and $-10$ \kms\ (Sembach, Howk, \& Savage
2000).  Figure \ref{fig:nav} shows that the B/O ratio changes between
these two regions.  The integrated abundances in these two components
are significantly different: $[{\rm B/O}]=-0.57\err{0.12}{0.16}$ for
the component centered at $v_{\rm LSR} = 0$ \kms, and $[{\rm B/O}] =
+0.08\err{0.13}{0.17}$ for the blend of warm components at negative
velocities.  This sightline exhibits variations in [B/O] that are
coupled to real changes in the physical properties of the observed
components.

The dependence of [B/O] on \avedens\ could potentially be caused by
other effects, including true abundance variations (gas+dust) and
differential ionization (e.g., Sembach et al. 2000).  Boron may be
particularly sensitive to the latter effect since its second
ionization potential is high (25.15 eV, similar to that of
\protect\ion{C}{2}).  While the effects of differential ionization 
may modify the component-to-component B/O ratios, they are likely not
large enough (perhaps $\la 0.1$ dex) to cause the 0.8 dex range of
[B/O] seen in Figure \ref{fig:b2o}.

We believe the incorporation of boron into grains is dominant among
the possible effects leading to the trend seen in Figure
\ref{fig:b2o}.  It is reasonable to expect that boron should be
incorporated into dust.  It has a condensation temperature ($910-964$
K; Zhai 1995; Lauretta \& Lodders 1996) similar to those of gallium
and copper and is in the same group of the periodic table as aluminum
and gallium, all of which are known to be significantly incorporated
into interstellar dust (Hobbs et al. 1993; Savage \& Sembach 1996).

The cold cloud towards \hd 104705 (Figure \ref{fig:nav}) shows a solar
B/Ga ratio, though the blend of warm components exhibits super-solar
B/Ga ratios.  If the abundance variations between these regions are
caused by the destruction of dust (see Savage \& Sembach 1996), then
boron is more readily-stripped from grains than is gallium.

The gas-phase abundance measurements of boron in Table
\ref{tab:abundances} yield no firm information on the solid-phase 
abundance of boron.  We derive a lower limit to the present-day total
(gas+dust) interstellar boron abundance of ${\rm B/H} \ga
(2.5\pm0.9)\times \e{-10}$ (using the measured [B/O] towards \hd
121968 and assuming $[{\rm O/H}] = -0.34\pm0.02$ from Meyer et al.
1998 corrected for the Morton 2000 \OI\ $f$-value).  The interstellar
B/H is lower than the meteoritic abundances
(${\rm B/H} = [7.6\pm0.7] \times\e{-10}$; Anders \& Grevesse 1989) 
and non-LTE abundances for stars of solar-like metallicity (${\rm B/H}
= 5 \times \e{-10}$; see discussion in Lambert et al. 1998).

The probable incorporation of boron into interstellar dust makes it
difficult to use the measured gas-phase interstellar boron abundance
to study the influence of spallation on the chemical evolution of the
light elements.  Another probe of spallation-induced chemical
evolution is the $^{11}$B/$^{10}$B isotope ratio (cf., Lambert et
al. 1998), and some of the sightlines presented here might allow
accurate measurements of this ratio with higher resolution or signal
to noise.  In fact, the sightline towards \hd 104705 seems to show a
high $^{11}$B/$^{10}$B ratio; the red wing of \BII\ profile is very
similar to that of the \protect\ion{Ga}{2} profile.  However, the
relative component structure seen in the \nav\ profiles of \BII\
towards \hd 104705, \hd 177989, and \hd 218915 are significantly
different than all of the other ionic species covered by our
observations, including \protect\ion{Ga}{2},
\protect\ion{Cu}{2}, and \protect\ion{Ge}{2}.  The depletion and/or 
ionization characteristics of \BII\ are disimilar to these species,
perhaps making it inappropriate to use them as templates for \BII\
when studying the $^{11}$B/$^{10}$B ratio along complicated
sightlines.

\section{Summary}

We have presented new and archival observations of the gas-phase
abundance of boron in the diffuse interstellar medium using STIS and
GHRS.  From our analysis of these high-quality absorption line data,
and measurements from the literature, we have concluded the following.

\begin{enumerate}

\item The gas-phase abundance of [B/O] in the ISM is anticorrelated 
with the average density of hydrogen along the sightline being probed,
as well as the fraction of hydrogen seen in molecular form along a
sightline.  Along individual sightlines, we also find a significantly
higher gas-phase [B/O] abundance in warm than in cold diffuse clouds.
The evidence strongly suggests that boron is incorporated into dust
grains in the diffuse ISM.

\item The relative component-to-component strengths in the observed 
\BII\ profiles  are significantly different than those of any other 
observed species, including \ion{Ga}{2}, \ion{Cu}{2}, and \ion{Ge}{2}.
The depletion and/or ionization characteristics of \BII\ are different
than those of other species.  Although \ion{Ga}{2}, \ion{Cu}{2}, and
\ion{Ge}{2} are sometimes used as templates for modeling the \BII\
absorption, the differences seen in our data suggest it may be
inappropriate to assume they trace the \BII\ profile when deriving the
$^{11}$B/$^{10}$B ratio along complicated sightlines.

\item We derive a lower limit to the present-day total (gas+dust) B/H 
abundance of ${\rm B/H} \ga (2.5\pm0.9)\times \e{-10}$.

\end{enumerate}

\acknowledgements

We thank S. Federman for suggestions on this work.  This work was
supported by NASA through grants GO-0720.01-96A and GO-0720.02-96A
from the Space Telescope Science Institute, which is operated by the
Association of Universities for Research in Astronomy, Inc., under
NASA contract NAS5-26555.



\pagebreak

\begin{deluxetable}{lcccc}
\tablenum{1}
\tablecolumns{5}
\tablewidth{0pc}
\tablecaption{Equivalent Width Measurements
 	\label{tab:eqwidths}}
\tablehead{
\colhead{} & \multicolumn{2}{c}{$W_\lambda$ [m\AA]\tablenotemark{a}} \\
\cline{2-3}
\colhead{Star} & 
\colhead{\ion{O}{1} $\lambda1355$} & 
\colhead{\ion{B}{2} $\lambda1362$} &
\colhead{$(v_{-},v_{+})$\tablenotemark{b}} &
\colhead{S/N\tablenotemark{c}}
}
\startdata
\cutinhead{STIS}
HD 104705 & $11.8\pm0.8$ & $5.3\pm1.2$ & $(-48, +7)$  & 80, 60 \\
HD 121968 &  $3.0\pm0.5$ & $1.8\pm0.6$ & $(-12,+10)$ & 55, 50 \\
HD 177989 & $10.4\pm0.5$ & $4.8\pm0.4$ & $(-12,+14)$ & 70, 65 \\
HD 218915 & $14.3\pm0.9$ & $2.3\pm0.6$ & $(-17, +2)$  & 60, 70 \\
HD 303308 & $18.4\pm1.5$ & $8.0\pm1.6$ & $(-20,+21)$ & 45, 40 \\
\cutinhead{GHRS}
$\alpha$ Sco B & $11.4\pm0.8$ & $1.6\pm0.5$ & $(-5, +8)$ & 35, 35 \\
$\zeta$ Oph    & $7.3\pm0.8$\tablenotemark{d}
		  & $0.79\pm0.17$ & $(-21,+7)$ & 56, 370 \\
\enddata
\tablenotetext{a}{Measured equivalent width and $1\sigma$ uncertainties 
	(in m\AA).}
\tablenotetext{b}{The range in $v_{\rm LSR}$ over which the absorption 
	profiles were integrated.}
\tablenotetext{c}{Empirically estimated signal-to-noise ratios for 
	continuum regions near the \protect\ion{O}{1} and
	\protect\ion{B}{2} absorption lines, respectively.}
\tablenotetext{d}{The \protect\ion{O}{1} $\lambda1355$ measurement 
	for $\zeta$ Oph is from Savage, Cardelli, \& Sofia (1992) and
	represents the sum of their components A and B.}
\end{deluxetable}


\pagebreak

\renewcommand{\err}[2]{\ensuremath{^{+ #1}_{- #2}}}

\begin{deluxetable}{lcllllllc}
\tablenum{2}
\tablecolumns{9}
\tablewidth{0pc}
\tablecaption{STIS and GHRS Interstellar Boron Abundance Measurements
	\label{tab:abundances}}
\tablehead{
\colhead{Star} & 
\colhead{ID\tablenotemark{a}} &
\colhead{$l, b$ \tablenotemark{b}} & 
\colhead{$d_*$ [pc]\tablenotemark{c}} &
\colhead{$\log N(\mbox{\ion{O}{1}})$\tablenotemark{d}} & 
\colhead{$\log N(\mbox{\ion{B}{2}})$\tablenotemark{d}} &
\colhead{$\log N({\rm H})$\tablenotemark{e}} &
\colhead{[B/O]\tablenotemark{f}} & 
\colhead{Ref.\tablenotemark{g}} 
}
\startdata
HD 104705 & 1 & $297.4,-0.3$ & 3900 &
	$17.83\pm0.03$ & $11.53\err{0.09}{0.11}$ &
	$21.11\err{0.25}{0.07}$ & $-0.31\err{0.10}{0.12}$ & 1 \\
HD 121968 & 2 & $334.0, +55.8$ & 3600 & 
	$17.21\pm0.07$ & $11.06\err{0.12}{0.18}$ &
	$20.71\err{0.25}{0.08}$ & $-0.17\err{0.14}{0.19}$ &  1 \\	 
HD 177989 & 3 & \phn$17.8, -11.9$ & 4900 &
	$17.84\pm0.03$\tablenotemark{h} & $11.48\err{0.04}{0.04}$ &
	$20.95\err{0.25}{0.09}$ & $-0.37\err{0.05}{0.05}$ & 1 \\  
HD 218915 & 4 & $108.1, -6.9$ & 3600 &
	$17.97\pm0.03$\tablenotemark{h} & $11.16\err{0.10}{0.13}$ &
	$21.30\err{0.25}{0.08}$ & $-0.82\err{0.10}{0.13}$ &  1 \\  
HD 303308 & 5 & $287.6, -0.6$ & 2600 & 
	$18.10\pm0.04$\tablenotemark{h} & $11.71\err{0.08}{0.10}$ &
	$21.45\err{0.25}{0.09}$ & $-0.40\err{0.09}{0.11}$ &  1 \\  
$\zeta$ Oph    & 6  & \phn\phn$6.3, +23.6$ & 140\err{15}{12} &
	$17.68\pm0.05$ & $10.69\err{0.09}{0.11}$ &
	$21.15\err{0.03}{0.03}$ & $-1.00\err{0.10}{0.12}$ & 1,2 \\
$\alpha$ Sco B & 7  & $352.0, +15.1$ & 158 & 
	$17.91\pm0.03$\tablenotemark{h} & $11.01\err{0.12}{0.16}$ &
	$21.43\err{0.25}{0.10}$ & $-0.91\err{0.12}{0.16}$ & 1   \\
$\delta$ Sco   & 8 & $350.1, +22.5$  & 123\err{15}{12} &
	$17.74\pm0.06$\tablenotemark{i} & $10.93\err{0.09}{0.11}$ &
	$21.08\err{0.07}{0.07}$ & $-0.82\err{0.11}{0.13}$ & 3,4 \\
$\kappa$ Ori   & 9 & $214.5, -18.5$ & 220\err{50}{32} &
	$17.03\pm0.04$ & $10.63\err{0.11}{0.15}$ &
	$20.53\err{0.04}{0.04}$ & $-0.41\err{0.12}{0.15}$ & 3,4 \\
$\lambda$ Ori  & 10 & $195.1, -12.0$ & 320\err{90}{60} &
	$17.33\pm0.06$ & $10.79\err{0.11}{0.15}$ &
	$20.81\err{0.07}{0.07}$ & $-0.55\err{0.13}{0.16}$ & 3,5 \\
$\iota$ Ori    & 11 & $209.5, -19.6$ & 410\err{180}{100} & 
	$16.76\pm0.07$ & $<10.30$ &
	$20.18\err{0.05}{0.05}$ & $<-0.47$ & 3,5 \\
\enddata
\tablenotetext{a}{Identification number designating  the observed stars 
	in Figure 2.}
\tablenotetext{b}{Galactic coordinates for the observed stars.}
\tablenotetext{c}{Distances to the observed stars.  Where available we
	have used {\it Hipparcos} (Perryman et al. 1998) distances
	(those distances with error bars).  Otherwise we have adopted
	the distances from Diplas \& Savage (1994).}
\tablenotetext{d}{Logarithmic column densities and $1\sigma$ error estimates 
	(in units atoms cm$^{-2}$) for \protect\ion{O}{1} and
	\protect\ion{B}{2} derived by integrating the apparent column
	density profiles.  We adopt oscillator strengths from Morton
	(2000): $f(\mbox{\ion{O}{1} }1355)= 1.16\times10^{-6}$ and
	$f(\mbox{\ion{B}{2} }1362)=0.998$.}
\tablenotetext{e}{Total logarithmic hydrogen column density 
	$N({\rm H}) = N(\mbox{\ion{H}{1}}) + 2N({\rm H}_2)$.  For many
	sightlines no H$_2$ column densities are available, and we
	have adopted \protect\ion{H}{1} column densities from Diplas
	\& Savage (1994) with an upper error bar of +0.25 dex.  All
	other measurements are a weighted mean of the Bohlin et
	al. (1978) and Diplas \& Savage (1994) \protect\ion{H}{1}
	column densities combined with the Savage et al. (1977) H$_2$
	column densities.}
\tablenotetext{f}{Relative logarithmic abundance of boron to oxygen,
	referenced to the relative solar system values such that
	$[{\rm B/O}] \equiv \log N(\mbox{\ion{B}{2}}) /
	N(\mbox{\ion{O}{1}}) - \log ({\rm B/O})_\odot$, where $\log
	({\rm B/O})_\odot = -5.99$ (Anders \& Grevesse 1989).}
\tablenotetext{g}{References for \protect\ion{O}{1} and 
	\protect\ion{B}{2} column densities: (1) This work; (2) Savage
	et al. 1992; (3) Meyer et al. 1998; (4) Lambert et al. 1998;
	(5) Jura et al. 1996.}
\tablenotetext{h}{Small saturation corrections have been made to
	the original apparent column densities based upon the
	\protect\ion{C}{1} curve of growth for each sightline.  The
	applied corrections are (in dex) +0.06, +0.06, +0.06, and
	+0.08 for HD~177989, HD~218915, HD~303308, and $\alpha$ Sco B,
	respectively.}
\tablenotetext{i}{The \protect\ion{O}{1} column density for the 
	$\delta$ Sco sightline is derived from {\em Copernicus}
	observations of the 1355 \AA\ transition (see Meyer et
	al. 1998).}
\end{deluxetable}

\pagebreak


\end{document}